\documentclass[aip, amsmath,amssymb,reprint,]{revtex4-1}

\usepackage{graphicx}% Include figure files
\usepackage{dcolumn}% Align table columns on decimal point
\usepackage{bm}% bold math
\usepackage[utf8]{inputenc}
\usepackage{graphicx}
\usepackage{subcaption}
\usepackage[T1]{fontenc}
\usepackage{mathptmx}
\usepackage{etoolbox}
\usepackage{hyperref}
\usepackage{lineno} 
\usepackage{comment}
\makeatletter
\def\@email#1#2{%
 \endgroup
 \patchcmd{\titleblock@produce}
  {\frontmatter@RRAPformat}
  {\frontmatter@RRAPformat{\produce@RRAP{*#1\href{mailto:#2}{#2}}}\frontmatter@RRAPformat}
  {}{}
}%

\makeatother
\begin{document}

\preprint{AIP/123-QED}

%\title{On Coherent Scattering and Imaging with Structured X-ray Beams}

\title{Manipulating Metastability: Quenched Control of Topological Defects in Multiferroics} 

% Force line breaks with \\
\author{Nimish P.~Nazirkar$^*$}
\affiliation{ 
Rensselaer Polytechnic Institute, Department of Materials Science and Engineering, Troy, NY, USA%\\This line break forced with \textbackslash\textbackslash
}%
\author{Sowmya Srinivasan}
\affiliation{ 
Rensselaer Polytechnic Institute, Department of Materials Science and Engineering, Troy, NY, USA%\\This line break forced with \textbackslash\textbackslash
}
%\author{Jian Shi}%
\author{Ross Harder}%
% \email{Second.Author@institution.edu.}
\affiliation{ 
Advanced Photon Source, Argonne National Lab, Lemont, Illinois, USA%\\This line break forced with \textbackslash\textbackslash
}%

\author{Edwin Fohtung}
 %\homepage{http://www.Second.institution.edu/~Charlie.Author.}
  \homepage{http://www.efohtung-research.com}
\affiliation{ Rensselaer Polytechnic Institute, Department of Materials Science and Engineering, Troy, NY, USA}
%\affiliation{Center for Materials, Devices, and Integrated Systems, Rensselaer Polytechnic Institute, Troy,NY, USA}

\email{nazirn@rpi.edu.}
\date{\today}% 

\begin{abstract}
The topological properties of quasiparticles, such as skyrmions and vortices, have the potential to offer extraordinary metastability through topological protection, and drive motion with minimal electrical current excitation. This has promising implications for future applications in spintronics. Skyrmions frequently appear either in lattice form or as separate, isolated quasiparticles \cite{Tokura21}. Magnetic ferroelectrics, a subset of multiferroics that exhibit magnetically induced ferroelectricity, possess intriguing characteristics like magnetic (electric) field-controlled ferroelectric (magnetic) responses. Previous research based on Landau theory indicated the potential to stabilize metastable phases in multiferroic barium hexaferrite \cite{Karpov19}. We have successfully stabilized these meta-stable phases through magnetic quenching of hexaferrite nanoparticles, leading to the creation of compelling topological structures. The structural changes in individual BaFe$_{12}$O$_{19}$ nanocrystals were scrutinized using Bragg coherent diffractive imaging, granting us insight into the emergent topological structures in field-quenched multiferroics. Additionally, we explored why these structures are energetically preferable for the formation of metastable topological structures \cite{Karpov2017,Karpov19}.

\end{abstract}

\maketitle

\section{Introduction}
Multiferroics, including hexagonal manganites and hexagonal ferrites, have garnered substantial attention in recent research due to their unique and diverse properties\cite{subodh95, Biplab95, adrian94, Nguyen98}. The presence of topological defects in these materials exhibit physical properties at the nanoscale that differ significantly from their bulk characteristics\cite{shi2023enhanced}. These properties, notably the coupling between structure, charge, and spin degrees of freedom, hold immense potential for revolutionizing the fields of electronics, energy storage, and material science\cite{Karpov19}. Given their exceptional magnetic properties and ability to facilitate controlled writing of topological defects, multiferroics are of significant interest.

\begin{figure*}[htb!]
\centering
\includegraphics[width=1\linewidth]{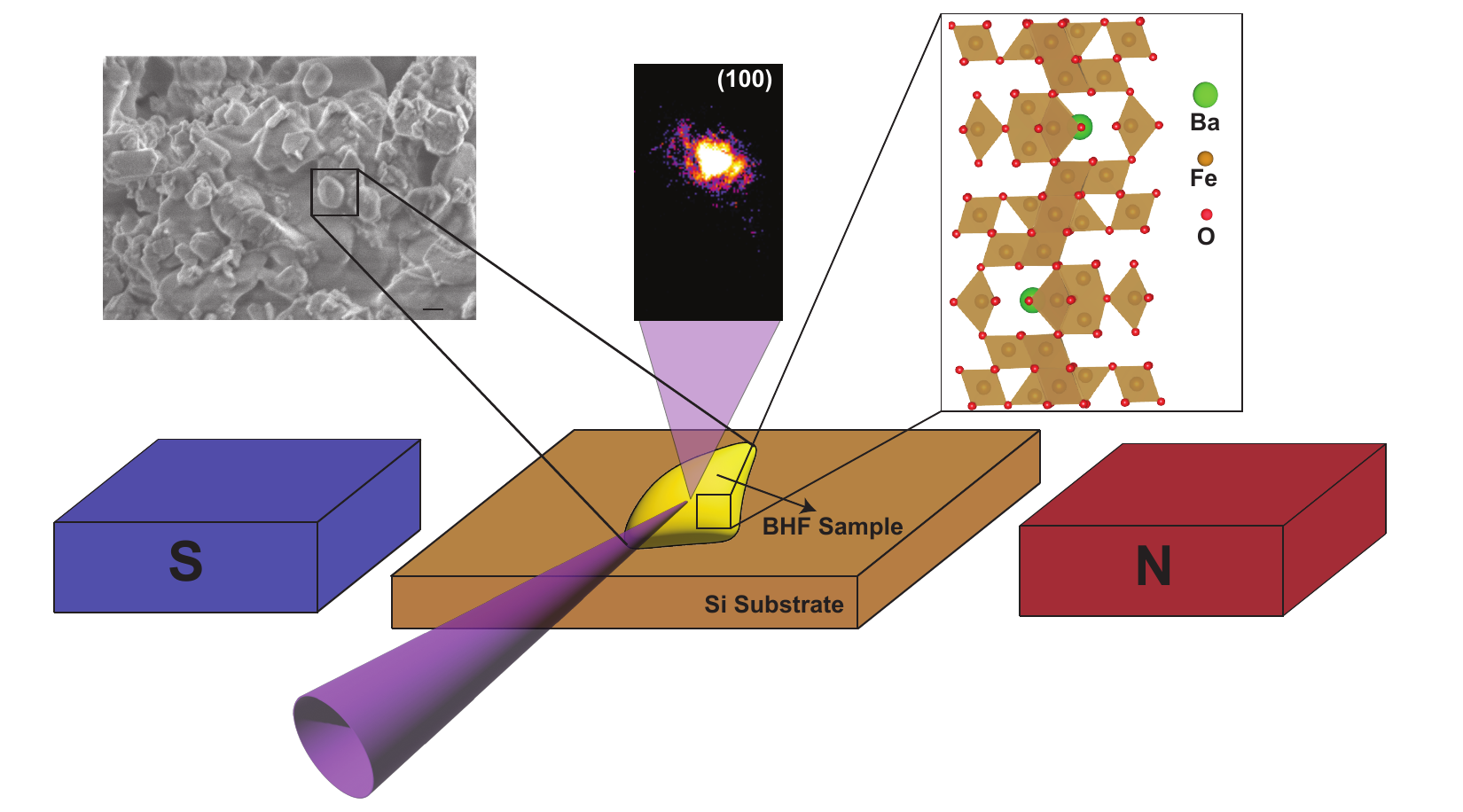}
\caption{\textbf{Crystal Structure and In-Operando X-ray Bragg Coherent Diffraction Experiment on $BaFe_{12}O_{19}$}. This figure showcases a Scanning Electron Microscopy (SEM) image of a sintered $BaFe_{12}O_{19}$ (BHF) sample. Coherent x-rays(depicted as purple cone) at 9.0 KeV were utilized to collect 3D Bragg diffraction(image at the center) near the (100) reciprocal lattice point (RLP) for an individual BHF nanoparticle in the presence of an external magnetic field. The scale bar in the SEM image represents 1$\mu$m.}
\label{fig1}
\end{figure*}

M-type Barium Hexa-ferrite (BHF) has emerged as a focal point of exploration as a permanent magnet, owing to its notable magnetization of $20 \mu B$ per formula unit\cite{Rowley2016,Kostishyn2016}. Furthermore, studies have unearthed BHF's fascinating dual nature as a quantum paraelectric at temperatures below 6K and as a bulk ferroelectric at room temperature, exhibiting a remarkable maximum remnant polarization of $11.8 \mu C/cm^2$ \cite{TAN201387}. This positions BHF as a promising candidate for a lead-free multiferroic. These two properties make BHF(Fig \ref{fig1}) a magnetic ferroelectric material where magnetism and ferroelectricity coexist, unlike other ferroelectric perovskites \cite{Hill2000}. Traditional multiferroics typically display ferroelectric and antiferromagnetic behavior. Nevertheless, these conventional systems do not exhibit standard indicators of ferroelectric instability such as large Born Effective charges\cite{}. For example, the spontaneous polarization ($P_s$) of $7.8 \mu C/cm^2$ in $InMnO_3$, resulting from the tilting of the Mn-O octahedra, suggests a phonon-mediated ferroelectric transition. This form of ferroelectricity is theorized to be controllable via magnetic fields\cite{bekheet2016}. However, the structural rationale behind this ferroelectric control and the behavior of the topological charges under a magnetic field remains elusive.

Introduced approximately two decades ago, Coherent X-ray Diffraction Imaging (CDI) \cite{miao1999extending} is a lensless imaging technique that captures the diffraction pattern produced by a coherent X-ray beam scattered by a sample. Unlike traditional methods which rely on lenses to form an image, CDI computationally reconstructs the sample's image from the measured diffracted wavefront, bypassing the limitations and aberrations introduced by lenses.

A specialized variant of CDI is Bragg Coherent Diffraction Imaging (BCDI), which zeroes in on the diffraction from crystalline samples satisfying Bragg's law. This specific approach has been extensively applied across various scientific fields. It has enabled researchers to investigate defects within crystals \cite{ulvestad2015topological,hofmann2020nanoscale,shi2020nanoscale,barringer2021imaging}, ferroelectric and ferroelastic domains \cite{diao2020evolution,Karpov2017,liu2020needle,shi2022topological}, and the nuanced dynamics of phase transitions. Furthermore, BCDI has provided deep insights into the internal strains of nanostructures, including but not limited to, nanoparticles and nanowires \cite{FohtungPRB2011,robinson2009coherent,hofmann2020nanoscale,shi2023enhanced}. This technique's adaptability has also been showcased in its capacity to study how materials respond to varied conditions, such as electrocatalytic reactions \cite{vicente2021bragg}, mechanical stress \cite{liu2020needle}, temperature variations \cite{yang2021annealing}, and changes in magnetic fields \cite{pateras2019room}.

The essence of BCDI, and by extension CDI, is underpinned by its advanced phase retrieval algorithms. These algorithms play a crucial role in converting coherent diffraction intensities in reciprocal space into a real-space complex wave function, all the while abiding by a rigorous oversampling criterion. Phase retrieval from coherent X-ray diffraction patterns is intricate, posing challenges that are fundamental to an array of disciplines including physics, materials science, and imaging. In this context, it's noteworthy that while the intensity of the diffracted X-rays can be measured directly, the phase information is elusive. Yet, the meticulous reconstruction of this phase data is imperative for generating the real-space image of the sample.

In this study, we employ an in-operando Bragg coherent diffraction imaging experiment (Fig \ref{fig1}) to examine the structural changes in BHF nanoparticles when subjected to an external field and subsequently quenched under said field. Our observations reveal the ferroelectric landscape transitioning between states under the influence of external fields and the quenching of the nanoparticle within these fields. The findings indicate the presence of a topological defect structure in the system in the absence of fields. Remarkably, these topological defects seem to organize into a lattice-like structure upon quenching in the field. The emergence of these defect structures is aligned with the $Z_2 x Z_3$ ferroelectric landscape, corroborating the predictions made by the Landau Theory \cite{Karpov19}. Our results suggest that these topological defects can be harnessed as a controlled-order parameter for ferroelectric materials.

\section{Results}

To perform BCDI on BHF nanoparticles, as shown in Fig~\ref{fig1}, the particles were first synthesized from a polymer precursor. Following synthesis, they were sintered into pellets \cite{Karpov19}. Afterward, the particles were placed on a Si(004) substrate in a randomized orientation. This allowed for the isolation of diffraction peaks during the BCDI experiment. During the experiment, magnetic fields were introduced using magnets adjacent to the sample, as depicted in Fig~\ref{fig1}.

\begin{figure*}[htb!] 
	\centering
	\includegraphics[width=1\linewidth]{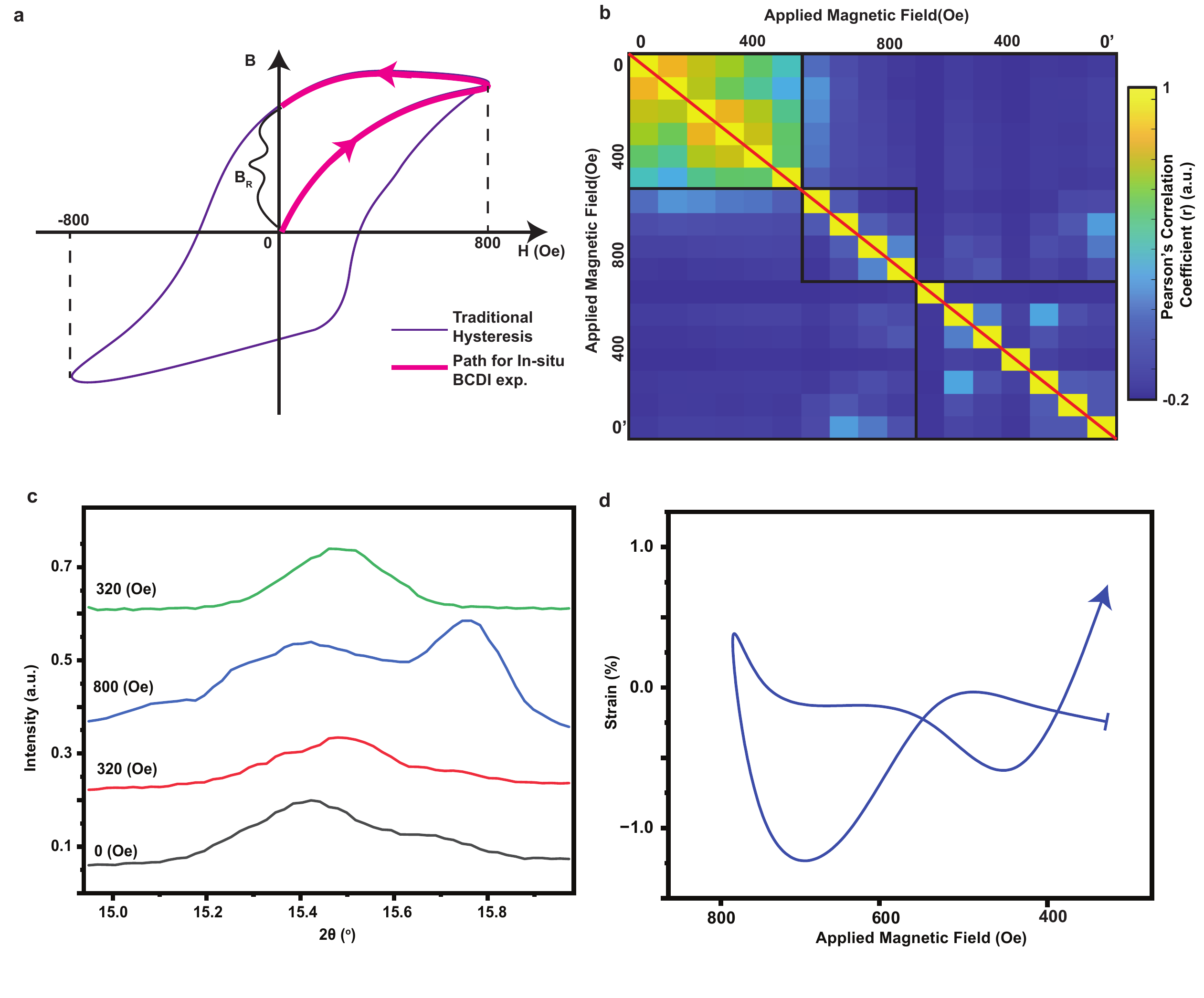}
	\caption{\textbf{Correlation of Strain Dynamics During Field Quenching:} \textbf{a:} Schematic of a traditional hysteresis loop for a magnetic material, illustrating the path followed for external magnetic field quenching in this experiment. The field is applied from $0 \, \text{Oe}$ to a maximum of $800 \, \text{Oe}$ and then quenched back to $0 \, \text{Oe}$. \textbf{b:} Pearson Correlation Coefficient map between CXD patterns as a function of an external magnetic field, indicating an irreversible structural transition and remnant strain in the nanoparticle upon quenching in the magnetic field. \textbf{c:} Line plots depicting the changes in the Bragg peak as a function of the applied field, providing evidence of inhomogeneous strain induced by the magnetic field. \textbf{d:} Asymmetric butterfly loop of strain response as a function of the external magnetic field, exhibiting characteristics of magnetostriction and electromechanical coupling associated with piezoelectric response in ferroelectrics.}
	\label{fig2}
\end{figure*}

We observed the (100) diffraction peak for changes in structural phase transitions in this system. A Pearson correlation of the diffraction patterns under varied magnetic fields was conducted to study the structural changes in the BHF nanoparticle (Fig \ref{fig2}\textbf{a}). We noticed an irreversible structural transition when cycled in a magnetic field of about 800 Oe, indicative of a structurally driven electronic transition through the application of magnetic fields. Inhomogeneity in the Bragg Peaks was evident, signifying a shift from the Bragg position and peak broadening (see Fig \ref{fig2}\textbf{b}). The structural strain observed in the particle under a magnetic field is depicted in Fig \ref{fig2}\textbf{c}. The strain exhibited the anticipated behavior of ferroelectric material under an electric field, providing direct evidence of magnetically driven ferroelectricity in multiferroic materials.

\begin{figure*}[htb!] 
	\centering
	\includegraphics[width=1\linewidth]{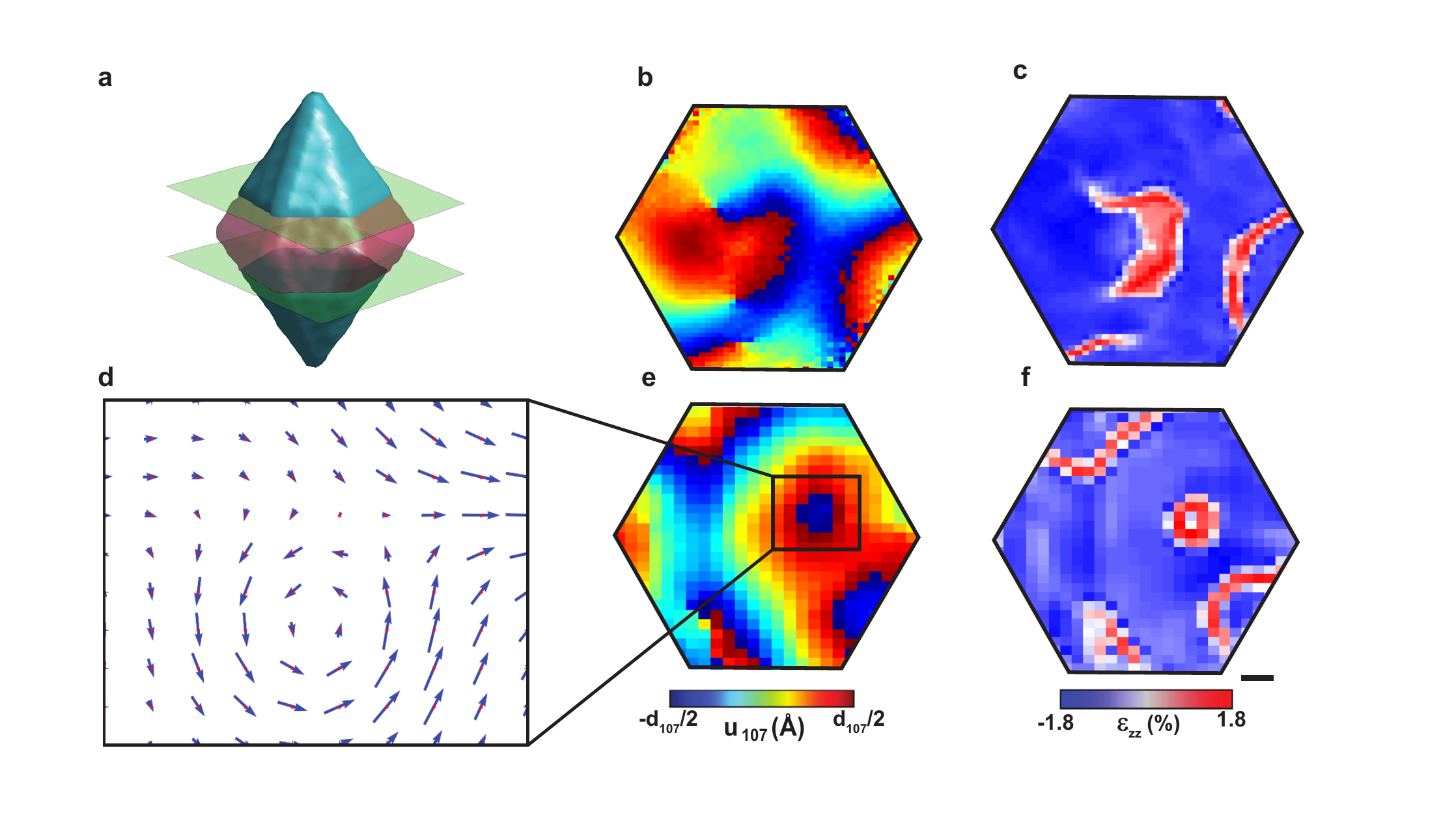}
	\caption{\textbf{Reconstruction Analysis:} \textbf{a:} 3D rendering of the Bragg Electronic density isosurface obtained from the reconstructions\cite{Karpov19}. \textbf{b:} Reconstructed displacement field of the nanoparticle under no applied field at the central slice. \textbf{c:} Strain in the particle under no applied field obtained from the reconstruction of the central slice. \textbf{d:} Dipole moment around the topological defect structure after the particle is quenched under an external field. \textbf{e:} Central Slice of the Reconstructed displacement field of the nanoparticle quenched under an external field. \textbf{f:} Central slice depicting the strain in the particle quenched under an applied field obtained from the reconstruction. The scale bar for \textbf{a}, \textbf{b}, \textbf{c}, \textbf{e}, and \textbf{f} is 50nm. }
	\label{fig3}
\end{figure*}

\begin{figure*}[htb!] 
	\centering
	\includegraphics[width=1\linewidth]{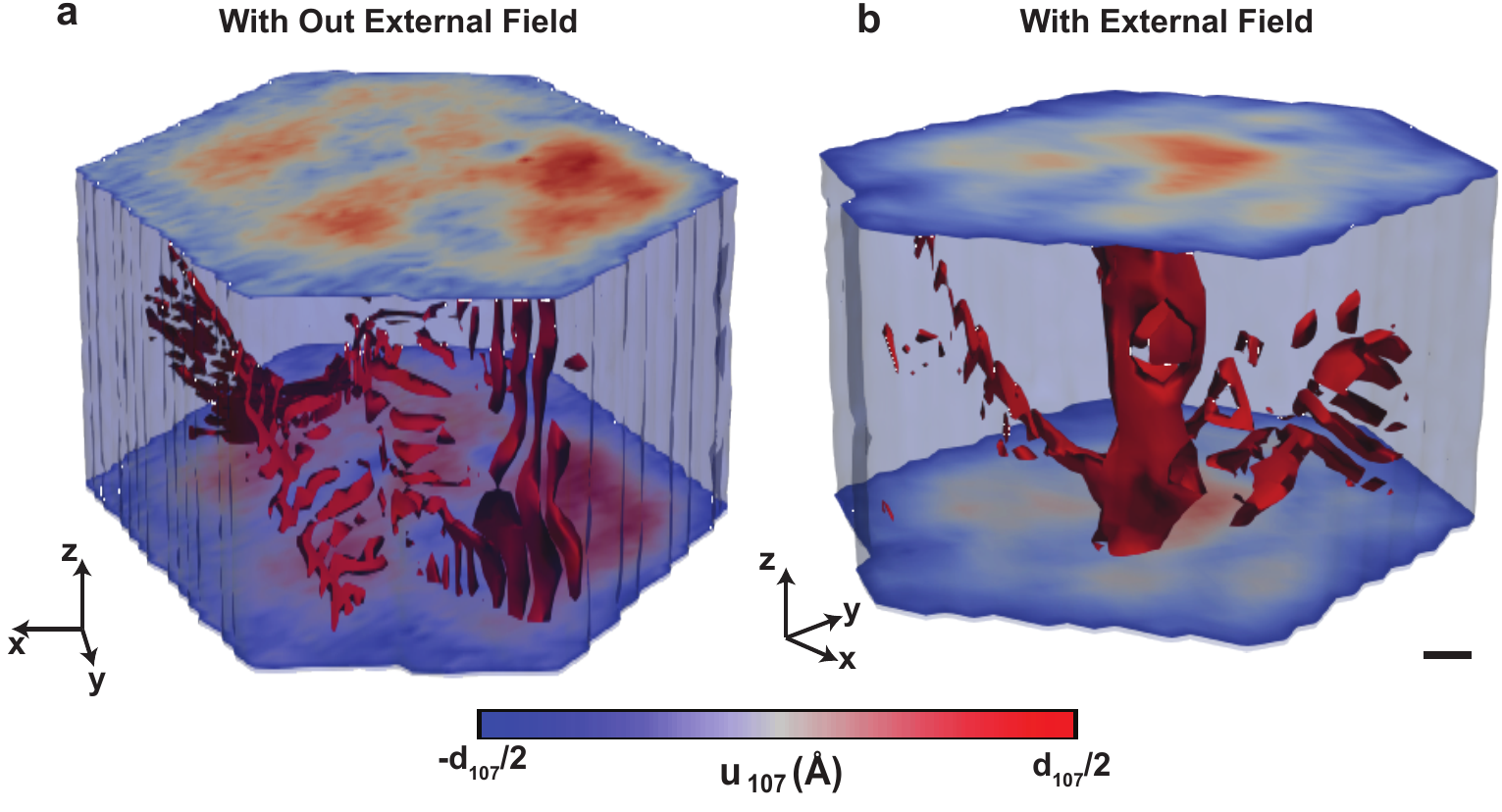}
	\caption{\textbf{Transparent Density Isosurface Tracking of Topological Structures.} 
	(a) Isosurface showing a network of topological defects in the nanocrystal without any external field. 
	(b) Isosurface of the quenched topological defects under a field, revealing a central lattice-like structure. 
	Coloring indicates lattice displacements as shown in fig.\ref{fig3}. The scale bar for both (a) and (b) is 50nm.}
	\label{fig4}
\end{figure*}

To gather the 3D displacement and strain fields of the nanoparticles, we rocked the crystal around the (100) Bragg peak. This approach aided in understanding and tracking the structural phase transition in the asymmetric diffraction peaks. Subsequently, iterative phase retrieval algorithms were employed to transition from diffraction space to 3D strain. The magnitude of the 3D Bragg Electronic density of BHF particles (Fig \ref{fig3} \textbf{a}) was analyzed to reveal the shape and size of the nanoparticle. The resolution limit of the reconstructed Bragg electronic density is approximately 15 nm. The imaginary part of this complex electronic density displays features indicative of the internal displacement field, with the phase shift of this complex density correlating to about $1.4$ \text{\AA} displacement from the ideal crystal structure. This displacement is directly associated with the ferroelectric landscape in the material. The ferroelectric displacement field, in the absence of any external field, demonstrates the existence of pairs of vortex and anti-vortex in the measured nanoparticle (Fig \ref{fig3} \textbf{b}). These pairs originate due to the presence of the $Z_2 x Z_3$ domain structure intersecting at points where the vortex or anti-vortex is observed.

However, after partial cycling or partial quenching under a field, it appears that the vortex and anti-vortex pairs coalesce to form a lattice-like structure, achieving complete flux closure as depicted in Fig \ref{fig3} \textbf{d} and \textbf{e}. Upon quenching in the applied external field, the domain structure in the nanoparticle reorganizes and stabilizes in a meta-stable ferroelectric phase. Strain is calculated by taking the gradient of the reconstructed displacement field. The observed strain without a field confirms the presence of different vortex and anti-vortex regions (see Fig \ref{fig3} \textbf{c}). When quenched under an external field, the strain validates the coalescence of the vortex and anti-vortex pairs to form a lattice-like structure, as seen in Fig \ref{fig3} \textbf{f}.

Further investigations into the 3D nature of the topological defects or strings reveal that, without any applied external field, a network of vortex loops is observable in the nanocrystal (Fig \ref{fig4} \textbf{a}). These topological defects form at the intersection of the ferroelectric domains, which are a result of the $Z_2 x Z_3$ symmetry in the ferroelectric landscape of the nanoparticle. This domain structure encompasses three major domains, each paired with an anti-domain, contributing to the $Z_2 x Z_3$ symmetry of the system. After quenching under an external field, a lattice-like structure of the topological defect is observed in the nanoparticle (see Fig \ref{fig4}\textbf{b}). The proposed mechanism is that during the quenching of the particle under an external field, the ferroelectric domains shift into a meta-stable structural phase, rearranging the topological defects into a lattice structure.

Consistent observations of inhomogeneous strain, combined with reconstructed and modeled Landau simulations, indicate a resemblance with certain hexa-manganites and traditional ferroelectric materials. Specifically, the strain behavior under an external magnetic field appears to be ferroelectric-like. This suggests that these topological defects are analogous to those observed in the aforementioned materials. Furthermore, these topological defects can be viewed as 3D quasi-particles. These can be manipulated by \textit{quenching} under external fields.

\section{Discussion}

\subsection{Strain relaxation due to quenching}

As shown in Fig \ref{fig2} \textbf{d} strain relaxation and irreversible structural transitions are observed in BHF nanoparticles due to quenching in an external magnetic field. To fully intepret our observation, strain relaxation in such scenarios necessitates a detailed model encompassing both mechanical and magnetic interactions.

One can use the theory of linear magnetoelasticity, which couples mechanical strain, \( \epsilon_{kl} \) , with the magnetic field, \( H \), by introducing magnetoelastic coupling constants. The total strain in the system, \( \epsilon_{\text{tot}} \) (See Table \ref{tab}), can be expressed as the sum of the elastic strain, \( \epsilon_{\text{el}} \), the thermal strain, \( \epsilon_{\text{th}} \), and the magnetoelastic strain, \( \epsilon_{\text{me}} \):

\[ \epsilon_{\text{tot}} = \epsilon_{\text{el}} + \epsilon_{\text{th}} + \epsilon_{\text{me}} \]

The magnetoelastic strain, \( \epsilon_{\text{me}} \), can be related to the magnetic field by the magnetoelastic constants. This strain component manifests due to the interaction of the magnetic field with the material, leading to a rearrangement of the domain structure and resulting in an irreversible structural transition. The magnitude and direction of the applied magnetic field during quenching play a crucial role in determining the nature of this transition and the resultant strain relaxation.

For a system like BHF nanoparticles exhibiting such irreversible transitions, advanced models such as Landau theories or phase-field models may be essential. These models can effectively describe the evolution of magnetic domains, the formation of topological defects, and the complex interplay between mechanical and magnetic states in multiferroic materials.

By solving these coupled equations, and incorporating the appropriate boundary conditions and constitutive relations, one can theoretically predict the strain relaxation and structural transitions induced by magnetic field quenching. Validation of these theoretical models would ideally be conducted by comparing their predictions with the experimental observations, such as those obtained through Bragg Coherent Diffraction Imaging (BCDI), to gain deeper insights into the strain dynamics and structural modifications in multiferroic nanoparticles.

\begin{table}[ht]
\centering
\small
\begin{tabular}{|l|p{2cm}|p{2.5cm}|p{2.5cm}|}
\hline
\textbf{Condition} & \parbox{2cm}{\textbf{Total} \\ \textbf{Strain(\%)}} & \parbox{2.5cm}{\textbf{Central} \\ \textbf{Strain(\%)}} & \parbox{2.5cm}{\textbf{Surface} \\ \textbf{Strain(\%)}}
\\
\hline
Unquenched & 0.4 & 1.5 & 1.3\\
Quenched & 1.0 & 2.0 & 1.8 \\
\hline
\end{tabular}
\caption{\label{tab}Comparison of Strain in field quenched and unquenched nanoparticle.}
\end{table}

\subsection{Significance of an Asymmetric Butterfly Magnetic Field vs Strain Loop}

An asymmetric butterfly loop in a magnetic field vs strain plot can exhibit several important implications, particularly in the realms of materials science and condensed matter physics:

\begin{enumerate}
    \item \textbf{Nonlinearity and Hysteresis:} The asymmetry typically indicates the presence of nonlinearity and hysteresis in the material’s response to an applied magnetic field. This is crucial for applications such as magnetic memory devices, where hysteresis behavior is exploited to store information.
    
    \item \textbf{Magnetostriction:} The asymmetric loop is suggestive of magnetostrictive effects, where the material experiences a strain or deformation in response to the magnetic field. This is particularly significant in designing sensors, actuators, and transducers.
    
    \item \textbf{Energy Dissipation:} The asymmetric behavior could signify that there is energy dissipation in the system when the magnetic field is applied and removed. This is essential for assessing the energy efficiency of the material in practical applications.
    
    \item \textbf{Domain Wall Motion:} The asymmetric behavior could be indicative of complex domain wall dynamics, such as the pinning and depinning of magnetic domains. This can impact the magnetic properties of the material and is relevant for magnetic storage and spintronic devices.
    
    \item \textbf{Phase Transitions:} The asymmetry might signal that the material undergoes structural phase transitions under the influence of a magnetic field, which could be irreversible. Understanding these transitions is vital for exploiting the material's properties in technological applications.
    
    \item \textbf{Material Defects and Inhomogeneities:} The asymmetry in the loop might arise due to defects, inhomogeneities, or anisotropies within the material, affecting its macroscopic properties and performance in devices.
\end{enumerate}

Understanding the origin and implications of the asymmetric butterfly loop is essential for optimizing the material’s properties for specific applications and could open up new avenues for research and technological development in multifunctional and smart materials.

Materials demonstrate notably divergent properties at the nanoscale compared to their bulk counterparts, opening avenues for leveraging these enhanced behaviors through external perturbations. In particular, multiferroic materials at the nanoscale offer intriguing prospects, as they can be steered into meta-stable structures using a variety of external stimuli, including magnetic and electric fields, as well as thermal and optical influences.

The potential applications of these findings are promising. Topological defects, which can be actively controlled by external perturbations, serve as promising candidates for active components in nanoelectronics \cite{Artyukhin2014}. In this study, we concentrated on BHF nanoparticles, presenting a novel mechanism underpinned by experimental evidence for rearranging the improper ferroelectric domains within this system. Employing in-operando BCDI on a 500 nm nanoparticle, we delved into deciphering the electronic and structural mechanisms responsible for the formation of a topological defect lattice when quenched under an external field.

This research delineates both the control and imaging of topological defects in the presence of external fields and propounds a new paradigm in order parameter space for modifications induced by external fields in device architectures. The advancements in synchrotron sources pave the way for a more in-depth exploration into manipulating topological defects. Looking ahead, experiments involving ferroelectric capacitors and devices composed of nanocrystal arrays with inherent topological defects will be pivotal. Utilizing external fields to influence these defects and leveraging coherent scattering and BCDI methodologies will yield valuable insights into the dynamics of topological defects in a plethora of functional materials.

\begin{acknowledgments}

N. N. and E.F. acknowledge support from the US Department of Energy (DOE), Office of Science, under grant No. DE-SC0023148. E.F. also acknowledges support from the US Department of Defense, Air Force Office of Scientific Research (AFOSR), under award No. FA9550-23-1-0325 (Program Manager: Dr. Ali Sayir) for work on probing topological vortices and piezoelectric enhancements. This research used resources of the Advanced Photon Source (APS), a U.S. Department of Energy (DOE) Office of Science User Facility, operated for the DOE Office of Science by Argonne National Laboratory (ANL) under contract No. DE-AC02-06CH11357.

%\dots.
\end{acknowledgments}

\section*{Data Availability Statement}

The data that support the findings of this study are available from the corresponding author upon request.

\nocite{*}
%\bibliography{aipsamp}% Produces the bibliography via BibTeX.
\section{Reference}
\bibliography{aipsamp}% Produces the bibliography via BibTeX.

\end{document}